# Electron Acceleration via Trapping inside Ion Mirror-mode Structures within A Large-scale Magnetic Flux Rope

Z. H. Zhong[1,2], H. Zhang[1,3], M. Zhou[1,2]*, D. B. Graham[4], R. X. Tang[1,2], X. H. Deng[1,2], Yu. V. Khotyaintsev[4]

[1]Institute of Space Science and Technology, Nanchang University, Nanchang 330031, China

[2]School of Physics and Materials Science, Nanchang University, Nanchang, China

[3]Institute of Space Weather, Nanjing University of Information Science and Technology, Nanjing, China

[4]Swedish Institute of Space Physics, Uppsala, Sweden

*Corresponding author: monmentum82@gmail.com

**Abstract**

Fermi acceleration is believed as a crucial process for the acceleration of energetic electrons within flux ropes (FRs) during magnetic reconnection. However, in finite-length FRs with a large core field, the finite contracting and the escaping of electrons along the axis can significantly limit the efficiency of Fermi acceleration. Using observations from the Magnetospheric Multiscale mission in the magnetotail, we demonstrate that magnetic mirror structures inside the FR can effectively prevent the escape of energetic electrons and overcome the limitation of finite contraction. Energetic electrons were produced and formed a power-law energy distribution in these mirror structures. By evaluating the acceleration rates, we show that these energetic electrons can be continuously accelerated within the mirror structures near the central region of the FR. These results unveil a novel mechanism that is universally applicable to electron acceleration within FRs in space, laboratory, and astrophysical plasmas.

Magnetic reconnection is a fundamental physical process in many relativistic/non-relativistic and collisionless/collisional plasma systems[1–3]. During magnetic reconnection, magnetic energy is efficiently converted to thermal and kinetic energy of plasma, which is associated with the macroscopic reconfiguration of magnetic field topology. Reconnection is responsible for many explosive phenomena in space, such as solar flares and substorms. Energetic electrons, which may carry a significant portion of the released magnetic energy during magnetic reconnection, are frequently observed in these explosive phenomena[4,5].

A long-standing question in magnetic reconnection is how energetic electrons are produced[6]. In situ observations of enhanced energetic electron fluxes in flux ropes (FRs) point out the importance of FRs in generating energetic electrons[7–13]. It has been suggested that Fermi acceleration due to FR contraction is the dominant mechanism for producing energetic electrons in multi-FR reconnection[14,15]. However, Zhou et al.[16] find that $E_\parallel$ may be more important than Fermi acceleration in ion-scale FRs. The efficiency of Fermi acceleration in two-dimensional (2-D) simulations is limited due to electron trapping in non-accelerating region around the FR center, where the magnetic field lines stop shrinking[17]. Recently, three-dimensional (3-D) simulations[17–19] show that instabilities growing in 3-D, such as tearing and kink instabilities, could lead to field line chaos in and around FRs, allowing particle transport from the non-acceleration region to the acceleration region, producing a large number of energetic electrons through volume filling.

The Magnetospheric Multiscale (MMS) mission[20] provides quantitative evidence of electron acceleration within FRs by directly calculating the local electron acceleration rates from different mechanisms[13,21,22]. Notably, Zhong et al.[13] found that betatron acceleration was the dominant mechanism responsible for the production of > 100 keV electrons within an ion-scale FR, despite the comparable magnitudes of Fermi and betatron acceleration rates. This observation is consistent with the hypothesis that the field-aligned electrons move too fast along the field line and escape the acceleration region along the axial direction of the finite-length FR, hindering the efficiency of Fermi acceleration. Whereas, 3-D particle-in-cell (PIC) simulations usually employ periodic boundary condition in the out-of-plane direction, ignoring electron escape along the axis of FR. Here, we demonstrate that mirror structures within the FR can act as "electron catchers", trapping electrons within the acceleration region, which may lead to the production of energetic electrons.

**Results**

**Event Overview.** A large-scale FR was observed by MMS on May 28, 2017, when the spacecraft was located at [-19.3, -11.8, 0.8] $R_E$ (Earth radius) in the Geocentric Solar Magnetospheric (GSM) coordinates. Four MMS spacecraft, arranged in a tetrahedron formation with an average spacing of ~ 60 km, obtained very similar observations of this FR. This study utilized data collected by several instruments onboard MMS, including the Fluxgate Magnetometer (FGM)[23], the Fast Plasma Investigation (FPI)[24], the Electric Double Probes (EDP)[25,26], the Search Coil Magnetometer (SCM)[27], and the Fly's Eye Energetic Particle Spectrometer (FEEPS)[28,29].

Figures 1a-1f display the measurements by MMS1 during 03:47:00-04:03:00 UT on May 28, 2017. At ~ 03:49:30 UT, a high-speed tailward ion outflow was detected, which peaked at around $V_{iX} \approx$ -600 km/s at ~ 03:54:30 UT (Figure 1d). The reversal of the tailward ion bulk flow $V_{iX}$ to earthward at ~ 03:58:40 UT, coinciding with the change of $B_z$ (Figure 1a) from negative to positive, indicates that MMS observed a tailward retreating reconnection X-line[30–32]. The spacecraft traversed the southern-lobe (near 03:56:57 UT) and northern-lobe (near 04:00:05 UT) regions before and after crossing the reconnection X-line (marked by the red dashed lines in Figures 1a-1f), respectively. These lobe regions are characterized by a large magnitude of $B_x$, very low plasma density (Figure 1c), weak ion bulk flow, and small differential energy fluxes of ions and electrons (Figures 1e and 1f). Based on the magnitude and shear angle of magnetic fields in these two lobe regions, we estimate a guide field $B_g$ of ~ 4 nT ~ 0.17 $B_0$, where $B_0$ is asymptotic reconnecting magnetic field ~ 24 nT.

There is an FR (orange shadow region) embedded in the tailward outflow[30] as shown in Figures 1a-1f, which is manifested by $B_z$ reversal from positive to negative and significant enhancement of $B_y$ (Figure 1a) and |**B**| (Figure 1b). Figure 2a shows the sketched trajectory of the MMS across the FR and the X-line. The duration of this FR is about 5.5 minutes, from 03:50:00 UT to 03:55:30 UT.

**Ion Mirror-mode Structures and Electrons Trapping.** Zoomed-in views of MMS1 observations during the 03:50:20-03:54:20 UT interval are presented in Figures 1g-1k. At the center of the FR (Figure 1g), a strong core field with a peak $B_y$ of ~ 41.5 nT was observed. Its direction is consistent with the guide field, while its magnitude is roughly 10 times the guide field. This extremely intense core field may be produced by the strong contraction and compression of the FR[33]. We note that both the total magnetic field |**B**| and the core field $B_y$ dramatically decrease to ~ 10 nT (about 25% of the maximal $B_y$) near and after the $B_z$ reversal point, between 03:52:40-03:53:45 UT (gray shaded region). Three magnetic cavities corresponding to enhancements in electron density (Figure 1h) were observed during this interval, which is similar to the previously reported magnetic cavity or hole[34].

The **X** component of the perpendicular electron (red curve) and ion (blue curve) bulk velocities agree well with the **E**×**B** drift velocity (black curve) in most regions of FR (see Figure 1i). This implies that the frozen-in condition is roughly satisfied during this interval. Thus, we use the average perpendicular electron bulk velocity in the **X** direction ~ -120 km/s to estimate the cross-sectional size of the FR, which is about 40,000 km ~ 6 $R_E$. In addition, we estimated that the magnetic cavities have a cross-sectional size of ~ 840-4,300 km ~ 2.8-14 $d_i$ (1 $d_i$ ~ 300 km is ion inertial length given the plasma density ~ 0.6 cm$^{-3}$).

Figure 1j shows the perpendicular ion temperature, $T_{i\perp}$ (blue curve), which increases in the FR and peaks within the magnetic cavities. The parallel ion temperature, $T_{i\|}$

(red curve), shows a minimum value at the center of the FR and increases as away from the center. While $T_{i\parallel}$ reaches several peaks during the interval when the magnetic cavities were observed, it is still smaller than $T_{i\perp}$. This ion temperature anisotropy satisfies the unstable criterion for the ion mirror instability (Figure 1k), $k = \frac{T_{i\perp}}{T_{i\parallel}} - \left(1 + \frac{1}{\beta_{i\perp}}\right) > 0$, where $\beta_{i\perp} = \frac{P_{i\perp}}{P_B}$ is the perpendicular ion plasma β. This indicates that the magnetic cavities are mirror structures generated by ion mirror instability[35–38].

Moreover, the principal axis (i.e., the maximal magnetic field variation direction) of these ion mirror structures is estimated using magnetic fields by the maximum variance analysis. The result suggests that the axis mainly lies along the L = [0.21, 0.97, 0.14] direction in GSM coordinates and is about 20° with respect to averaged background magnetic field. Additionally, we use the Maximum Directional Derivative (MDD)[39] method based on four-spacecraft magnetic fields to obtain a similar result of the principal axis, $L_{MDD}$ = [0.18, 0.85, -0.22], suggesting that the estimated principal axis is reliable. The principal axis mainly aligns with the axis of the FR (i.e., Y direction in GSM coordinates), as illustrated in Figure 2. Furthermore, we calculated the gradient of the magnetic field based on four-spacecraft measurements and found that the maximal gradient of |**B**| in the **Y** direction is $\nabla|\boldsymbol{B}|_{Ymax} \sim 0.01$ nT/km during this interval. By combining the magnitude change of the magnetic field, $\Delta|\boldsymbol{B}| = 31.5$ nT, from the mirror center (~ 10 nT) to the mirror point (~ 41.5 nT), we can roughly estimate the length of the mirror structures along its axis as $L = 2\frac{\Delta|\boldsymbol{B}|}{\nabla|\boldsymbol{B}|_{Ymax}} = $ 6,300 km ~ 1 $R_E$ ~ 21 $d_i$. Therefore, the axial length of the FR should be equal to or greater than 1 $R_E$.

Figure 3 presents observations of electrons in the vicinity of three magnetic mirror structures. Both parallel ($T_{e\parallel}$) and perpendicular ($T_{e\perp}$) electron temperatures increase within the mirror structures with respect to that in the leading part of FR (Figure 3b). Specifically, $T_{e\perp}$ is generally greater than $T_{e\parallel}$ at the edge of the mirror structures, while $T_{e\parallel}$ is generally greater than $T_{e\perp}$ at the central region of the mirrors (marked by grey shadows). Moreover, the fluxes of energetic electrons (≥ 47 keV) significantly increase and reach peaks within the mirror structures (Figure 3c). Figures 3d-3g display the pitch angle distributions (PADs) of electrons with different energies. The black curves in Figures 3d-3g represent the trapping-passing boundaries $\theta_{tr}$ and $180° - \theta_{tr}$, where $\theta_{tr} = sin^{-1}(\sqrt{|\boldsymbol{B}|/|\boldsymbol{B}_{max}|})$ is the trapping/loss angle of the charged particle in the mirror structure[40,41], and $B_{max}$ = 41.5 nT is the maximum magnetic field during this interval. Electrons with pitch angles between these two black curves are trapped inside the mirror structure and bounce between two mirror points, as illustrated in Figure 2b.

Figure 3d shows that low-energy electrons (30-1,000 eV) exhibit a cigar-shaped distribution within the mirror structures. The fluxes of these electrons decrease between the trapping-passing boundaries, possibly because they have been accelerated to higher energies. Figure 3e reveals that high-energy electrons (1-10 keV) are predominantly trapped by the mirror structures, while the fluxes of both parallel and anti-parallel electrons also increase at the center of the mirror structures, contributing to the peaks of $T_{e\parallel}$ observed in Figure 3b. Electrons with energies above 10 keV (Figures 3f and 3g) are mostly distributed between the trapping-passing boundaries, indicating that the magnetic field strength at the mirror points of these mirror structures is the same as the maximum magnetic field ($B_{max}$ = 41.5 nT) during this interval. MMS did not record any magnetic fields with strengths close to 41.5 nT between these mirror structures, which implies that the spacecraft did not cross the mirror points of these mirror structures. The energetic electrons (> 10 keV) are trapped near the central region of the FR by these magnetic mirror structures, which inhibits electron escaping along the axial direction of the FR. Although not all electrons are trapped, the trapping range of pitch angles will become broader as the mirror structures grow, corresponding to a smaller $|\mathbf{B}|/|\mathbf{B}_{max}|$. Figures 3h-3j display the electron energy spectra observed inside the three mirror structures, respectively, obtained by combining measurements from FPI and FEEPS. These energy spectra have the same power-law index (-4.6) in the energy range of 5-200 keV, consistent with that in reconnection outflow region in previous observations[42,43] and simulations[19]. The independence of the power-law index on time and location suggests that energetic electrons may be accelerated by a quasi-adiabatic process[42,44].

**Local Electron Acceleration Rates.** A crucial remaining question is whether the mirror structures coincide with the electron acceleration regions. If they are not, then the trapping effect of the mirror structures will inhibit the further acceleration of trapped electrons[18]. Conversely, this trapping effect will promote the persistent acceleration of trapped electrons, particularly for the Fermi mechanism, which accelerates charged particles in the field-aligned direction.

Therefore, we estimated the local electron acceleration rates under the guiding center approximation[45]. The local betatron acceleration rate can be calculated via[15,16,46]:

$$\partial_t W_b = P_{e\perp} \vec{v}_{E\times B} \cdot \frac{\nabla B}{B} + \frac{P_{e\perp}}{B} \frac{\partial B}{\partial t} \qquad (2)$$

and the local Fermi acceleration rate can be calculated via[15,16,46]:

$$\partial_t W_f = (P_{e\parallel} + n_e m_e v_\parallel^2) \vec{v}_{E\times B} \cdot (\vec{b} \cdot \nabla \vec{b}) \qquad (3)$$

where $n_e$ and $P_e$ represent the electron density and pressure, respectively. Furthermore, the local acceleration from parallel electric field can be calculated by[47,48]:

$$\partial_t W_{E_\parallel} = (J_{e\parallel} + \frac{\beta_{e\perp}}{2} J_\parallel) E_\parallel \qquad (4)$$

where $\beta_{e\perp} = P_{e\perp}/P_B$ is the perpendicular electron $\beta$ and $J_\parallel$ is the parallel current density.

Figure 4 shows the estimated local electron acceleration rates in the vicinity of the mirror structures. Mirror structure 1 is positioned at the center of the FR, coincident with the $B_z$ reversal point (denoted by a magenta dashed line). Mirror structures 2 and 3 are located at the trailing part ($B_z < 0$) of the FR (Figure 2b). Figure 4b shows the parallel electric fields $E_\parallel$ and the uncertainties measured by MMS1. One can see that $E_\parallel$ is typically smaller than the uncertainties, hence the electron acceleration from $E_\parallel$ can be neglected here.

Figure 4c displays the estimated Fermi acceleration rates $\partial_t W_f$. A significant negative peak ~ -150 eV/s·cm³ can be seen at the leading part ($B_z > 0$) of the FR in mirror 1, while it shows smaller fluctuated values in almost the other regions. To identify the acceleration region more clearly, we integrate $\partial_t W_f$ across this interval, $W_f = \int \partial_t W_f dt$, which is presented in Figure 4d. The negative and positive slopes of $W_f$ represent deceleration and acceleration, respectively. In the leading part ($B_z > 0$) of the FR, electrons experience deceleration by the Fermi mechanism, while the trailing part ($B_z < 0$) of the FR is the Fermi acceleration region, except for a small deceleration region at the center of mirror structure 3. This negative-positive variation of the Fermi acceleration rate across the FR is due to the tailward motion of the FR, which is consistent with previous observations and simulations[13,15,17,22]. Figures 4e and 4f show the betatron acceleration rate and its integration across this interval, respectively. Betatron acceleration region is mainly in the trailing part of mirror 1 and mirror 3, which corresponds to the total magnetic field enhancement, while the betatron deceleration region is located in the region of the total magnetic field decrease. These results suggest that Fermi and betatron acceleration mechanisms were still active in the central region of FR. Whereas, the Fermi electron acceleration mechanism appears to be the most important one, given that its positive acceleration rate occupies a larger area in the center of the FR.

**Discussion**

The estimated local electron acceleration rates indicate that magnetic mirror structures are in the Fermi acceleration region within the FR. Since these mirror structures are located almost in the trailing part ($B_z < 0$) of the FR and the mirror points in each flux tube distribute along the axis of the FR as illustrated in Figure 2b, they can effectively trap electrons and confine them in the Fermi acceleration region rather than allowing them to escape quickly along the axis of FR or move into the Fermi deceleration region (the leading part of the FR with $B_z > 0$) along the helical magnetic field lines. Another remarkable feature of these mirror structures is that they keep the Fermi acceleration active even if the moving FR stops contracting. Generally, a moving FR without contraction exhibits symmetric positive and negative Fermi acceleration rates on the two sides of the FR in the spacecraft frame[13,15,17]. Electrons streaming along the helical magnetic field lines of FR experience acceleration and deceleration alternatively and gain zero net energy eventually. However, mirror structures developed on the side with a positive Fermi acceleration rate can trap electrons in this acceleration region and lead to a net energy gain. This means that the mirror structures inside the FR provide a new pathway for electron energization in a moving and non-contracting FR, which is probably the final state of FR[49]. The Fermi acceleration can continue to increase the parallel energy of trapped electrons until they cross the trapping-passing boundaries and leave this region. Note that the magnetic mirror structures observed in this event were generated by ion mirror instability, however, similar mirror structures can be generated by many other mechanisms, such as electron mirror instability, Kelvin-Helmholtz instability, field-swelling instability, and solitary waves, which have been widely observed in various plasma environments[38,50–55]. Any mirror structures inside the FR can trap electrons, regardless of their generation mechanism.

Furthermore, intense plasma waves with a frequency between 0.1 $f_{ce}$ and 0.5 $f_{ce}$ were observed within the magnetic mirror structures (Figure 4g), consistent with previous observations[41,56]. The ellipticity of these waves is close to +1 (right-hand circular polarization, shown in Figure 4h), proving that they are electron whistler waves. The normalized parallel Poynting flux $S_\parallel/|S|$ displayed in Figure 4i illustrates that both parallel and anti-parallel propagating whistlers were observed, indicating that the spacecraft was crossing the source region of the whistlers[57]. These whistlers were generated by the perpendicular anisotropy of the trapped electrons[34,40,54], which may reduce the perpendicular energy of the trapped electrons and lead to fast pitch-angle scattering. This scattering could reduce the efficiency of the Fermi acceleration by freeing the energetic electrons from the mirror structures. The effect of whistlers on electron acceleration within FRs warrants further study.

In conclusion, MMS observed a large-scale FR with a cross-section size of about 6 $R_E$ in a reconnection outflow in the Earth's magnetotail. This FR has a large core field of ~ 41.5 nT, which is about 10 times the guide field (~ 4 nT) of the reconnection. We have studied the properties of the mirror structures and the local electron acceleration near the center of this FR. Our main results are summarized below:

1. Magnetic mirror structures generated by ion mirror instability are first observed in a large-scale FR in the magnetotail. The axes of these mirror structures are approximately aligned with the axial orientation of the FR.

2. The energetic electrons show a power-law energy distribution with an index of -4.6 within the magnetic mirror structures. These electrons were trapped by the mirror structures in the FR, which prevented them from escaping along the axial field of the FR.

3. The mirror structures are located in the electron acceleration region inside the FR, thereby facilitating the acceleration of the energetic electrons by Fermi mechanism and overcoming the limitation imposed by the finite contraction of the FR.

The mirror structures developed inside the acceleration region within the FR offer a novel scenario for electron energization inside reconnection-driven FRs. This scenario is different from the recent theory that electrons can easily access many acceleration regions along chaotic magnetic field lines resulting from the 3-D tearing or kink instabilities spontaneously formed inside/around FRs[17,19]. This new mechanism can be applicable to electron acceleration in a broad context where FRs have been frequently detected, such as planetary magnetosphere, solar atmosphere, etc.

**Acknowledgements**

We thank the entire MMS team and MMS Science Data Center for providing high-quality data for this study. This work was supported by the National Natural Science Foundation of China (NSFC) grants 42104156, 42130211, 42074197, and 41974195, the Jiangxi Provincial Natural Science Foundation grant 20224BAB211021, and the Project funded by China Postdoctoral Science Foundation grant 2021M691395.

**Data Availability**

The spacecraft data used for this study are publicly available from the MMS Science Data Center: https://lasp.colorado.edu/mms/sdc/public/about/browse-wrapper/.

**Author Contributions**

M.Z. conceived the idea of this study. Z.H.Z. and H.Z. carried out the data analysis. M.Z., Z.H.Z and H.Z. wrote the manuscript. D.B.G., R.X.T., X.H.D., and Y.V.K. participated in the interpretation of the data and the preparation of the manuscript. All the authors made significant contributions to this work.

**Reference**

1. Hesse, M. & Cassak, P. A. Magnetic Reconnection in the Space Sciences: Past, Present, and Future. *J. Geophys. Res. Space Phys.* **125**, (2020).

2. Liu, Y.-H. *et al.* First-principles theory of the rate of magnetic reconnection in magnetospheric

and solar plasmas. *Commun. Phys.* **5**, 97 (2022).

3. Zhao, Z. *et al.* Laboratory observation of plasmoid-dominated magnetic reconnection in hybrid collisional-collisionless regime. *Commun. Phys.* **5**, 247 (2022).

4. Holman, G. D., Sui, L., Schwartz, R. A. & Emslie, A. G. Electron Bremsstrahlung Hard X-Ray Spectra, Electron Distributions, and Energetics in the 2002 July 23 Solar Flare. *Astrophys. J.* **595**, L97–L101 (2003).

5. Lin, R. P. *et al. RHESSI* Observations of Particle Acceleration and Energy Release in an Intense Solar Gamma-Ray Line Flare. *Astrophys. J.* **595**, L69–L76 (2003).

6. Zhou, M. *et al.* Statistics of energetic electrons in the magnetotail reconnection. *J. Geophys. Res. Space Phys.* **121**, 3108–3119 (2016).

7. Zong, Q.-G. Cluster observations of earthward flowing plasmoid in the tail. *Geophys. Res. Lett.* **31**, L18803 (2004).

8. Chen, L.-J. *et al.* Observation of energetic electrons within magnetic islands. *Nat. Phys.* **4**, 19–23 (2008).

9. Retinò, A. *et al.* Cluster observations of energetic electrons and electromagnetic fields within a reconnecting thin current sheet in the Earth's magnetotail: ENERGETIC ELECTRONS IN CURRENT SHEET. *J. Geophys. Res. Space Phys.* **113**, n/a-n/a (2008).

10. Huang, S. Y. *et al.* Electron acceleration in the reconnection diffusion region: Cluster observations: ELECTRON ACCELERATION OBSERVATIONS. *Geophys. Res. Lett.* **39**, n/a-n/a (2012).

11. Wang, R., Lu, Q., Du, A. & Wang, S. *In Situ* Observations of a Secondary Magnetic Island in an Ion Diffusion Region and Associated Energetic Electrons. *Phys. Rev. Lett.* **104**, 175003

(2010).

12. Lu, S., Artemyev, A. V., Angelopoulos, V. & Pritchett, P. L. Energetic Electron Acceleration by Ion-scale Magnetic Islands in Turbulent Magnetic Reconnection: Particle-in-cell Simulations and ARTEMIS Observations. *Astrophys. J.* **896**, 105 (2020).

13. Zhong, Z. H. *et al.* Direct Evidence for Electron Acceleration Within Ion-Scale Flux Rope. *Geophys. Res. Lett.* **47**, (2020).

14. Drake, J. F., Swisdak, M., Che, H. & Shay, M. A. Electron acceleration from contracting magnetic islands during reconnection. *Nature* **443**, 553–556 (2006).

15. Dahlin, J. T., Drake, J. F. & Swisdak, M. The mechanisms of electron heating and acceleration during magnetic reconnection. *Phys. Plasmas* **21**, 092304 (2014).

16. Zhou, M. *et al.* Suprathermal Electron Acceleration in a Reconnecting Magnetotail: Large-Scale Kinetic Simulation. *J. Geophys. Res. Space Phys.* **123**, 8087–8108 (2018).

17. Dahlin, J. T., Drake, J. F. & Swisdak, M. The role of three-dimensional transport in driving enhanced electron acceleration during magnetic reconnection. *Phys. Plasmas* **24**, 092110 (2017).

18. Dahlin, J. T., Drake, J. F. & Swisdak, M. Electron acceleration in three-dimensional magnetic reconnection with a guide field. *Phys. Plasmas* **22**, 100704 (2015).

19. Zhang, Q., Guo, F., Daughton, W., Li, H. & Li, X. Efficient Nonthermal Ion and Electron Acceleration Enabled by the Flux-Rope Kink Instability in 3D Nonrelativistic Magnetic Reconnection. *Phys. Rev. Lett.* **127**, 185101 (2021).

20. Burch, J. L., Moore, T. E., Torbert, R. B. & Giles, B. L. Magnetospheric Multiscale Overview and Science Objectives. *Space Sci. Rev.* **199**, 5–21 (2016).

21. Akhavan-Tafti, M., Slavin, J. A., Sun, W. J., Le, G. & Gershman, D. J. MMS Observations of Plasma Heating Associated With FTE Growth. *Geophys. Res. Lett.* **46**, 12654–12664 (2019).

22. Jiang, K. *et al.* Statistical Properties of Current, Energy Conversion, and Electron Acceleration in Flux Ropes in the Terrestrial Magnetotail. *Geophys. Res. Lett.* **48**, (2021).

23. Russell, C. T. *et al.* The Magnetospheric Multiscale Magnetometers. *Space Sci. Rev.* **199**, 189–256 (2016).

24. Pollock, C. *et al.* Fast Plasma Investigation for Magnetospheric Multiscale. *Space Sci. Rev.* **199**, 331–406 (2016).

25. Ergun, R. E. *et al.* The Axial Double Probe and Fields Signal Processing for the MMS Mission. *Space Sci. Rev.* **199**, 167–188 (2016).

26. Lindqvist, P.-A. *et al.* The Spin-Plane Double Probe Electric Field Instrument for MMS. *Space Sci. Rev.* **199**, 137–165 (2016).

27. Le Contel, O. *et al.* The Search-Coil Magnetometer for MMS. *Space Sci. Rev.* **199**, 257–282 (2016).

28. Blake, J. B. *et al.* The Fly's Eye Energetic Particle Spectrometer (FEEPS) Sensors for the Magnetospheric Multiscale (MMS) Mission. *Space Sci. Rev.* **199**, 309–329 (2016).

29. Mauk, B. H. *et al.* The Energetic Particle Detector (EPD) Investigation and the Energetic Ion Spectrometer (EIS) for the Magnetospheric Multiscale (MMS) Mission. *Space Sci. Rev.* **199**, 471–514 (2016).

30. Zhou, M. *et al.* Observations of Secondary Magnetic Reconnection in the Turbulent Reconnection Outflow. *Geophys. Res. Lett.* **48**, (2021).

31. Jin, R., Zhou, M., Pang, Y., Deng, X. & Yi, Y. Characteristics of Turbulence Driven by Transient

Magnetic Reconnection in the Terrestrial Magnetotail. *Astrophys. J.* **925**, 17 (2022).

32. Li, X. *et al.* Three-dimensional network of filamentary currents and super-thermal electrons during magnetotail magnetic reconnection. *Nat. Commun.* **13**, 3241 (2022).

33. Lu, S. *et al.* Particle-in-cell Simulations of Secondary Magnetic Islands: Ion-scale Flux Ropes and Plasmoids. *Astrophys. J.* **900**, 145 (2020).

34. Zhang, H. *et al.* Modulation of Whistler Mode Waves by Ultra-Low Frequency Wave in a Macroscale Magnetic Hole: MMS Observations. *Geophys. Res. Lett.* **48**, (2021).

35. Hasegawa, A. Drift Mirror Instability in the Magnetosphere. *Phys. Fluids* **12**, 2642 (1969).

36. Ahmadi, N., Germaschewski, K. & Raeder, J. Simulation of magnetic holes formation in the magnetosheath. *Phys. Plasmas* **24**, 122121 (2017).

37. Zhang, L., He, J., Zhao, J., Yao, S. & Feng, X. Nature of Magnetic Holes above Ion Scales: A Mixture of Stable Slow Magnetosonic and Unstable Mirror Modes in a Double-polytropic Scenario? *Astrophys. J.* **864**, 35 (2018).

38. Zhong, Z. H. *et al.* Stacked Electron Diffusion Regions and Electron Kelvin–Helmholtz Vortices within the Ion Diffusion Region of Collisionless Magnetic Reconnection. *Astrophys. J. Lett.* **926**, L27 (2022).

39. Shi, Q. Q. *et al.* Dimensional analysis of observed structures using multipoint magnetic field measurements: Application to Cluster: STRUCTURE DIMENSIONALITY DETERMINATION. *Geophys. Res. Lett.* **32**, n/a-n/a (2005).

40. Breuillard, H. *et al.* The Properties of Lion Roars and Electron Dynamics in Mirror Mode Waves Observed by the Magnetospheric MultiScale Mission. *J. Geophys. Res. Space Phys.* **123**, 93–103 (2018).

41. Ahmadi, N. *et al.* Generation of Electron Whistler Waves at the Mirror Mode Magnetic Holes: MMS Observations and PIC Simulation. *J. Geophys. Res. Space Phys.* **123**, 6383–6393 (2018).

42. Fu, H. S., Khotyaintsev, Yu. V., Vaivads, A., Retinò, A. & André, M. Energetic electron acceleration by unsteady magnetic reconnection. *Nat. Phys.* **9**, 426–430 (2013).

43. Ergun, R. E. *et al.* Observations of Particle Acceleration in Magnetic Reconnection–driven Turbulence. *Astrophys. J.* **898**, 154 (2020).

44. Imada, S. *et al.* Energetic electron acceleration in the downstream reconnection outflow region: ENERGETIC ELECTRONS AND LARGE $B_z$. *J. Geophys. Res. Space Phys.* **112**, n/a-n/a (2007).

45. Northrop, T. G. Adiabatic charged-particle motion. *Rev. Geophys.* **1**, 283 (1963).

46. Dahlin, J. T., Drake, J. F. & Swisdak, M. Parallel electric fields are inefficient drivers of energetic electrons in magnetic reconnection. *Phys. Plasmas* **23**, 120704 (2016).

47. Ma, W., Zhou, M., Zhong, Z. & Deng, X. Electron Acceleration Rate at Dipolarization Fronts. *Astrophys. J.* **903**, 84 (2020).

48. Ma, W., Zhou, M., Zhong, Z. & Deng, X. Contrasting the Mechanisms of Reconnection-driven Electron Acceleration with In Situ Observations from MMS in the Terrestrial Magnetotail. *Astrophys. J.* **931**, 135 (2022).

49. Zhang, H. *et al.* Modeling a force-free flux transfer event probed by multiple Time History of Events and Macroscale Interactions during Substorms (THEMIS) spacecraft: MODELING A FORCE-FREE FTE. *J. Geophys. Res. Space Phys.* **113**, n/a-n/a (2008).

50. Russell, C. T., Riedler, W., Schwingenschuh, K. & Yeroshenko, Ye. Mirror instability in the magnetosphere of comet Halley. *Geophys. Res. Lett.* **14**, 644–647 (1987).

51. Huang, S. Y. *et al.* Magnetospheric Multiscale Observations of Electron Vortex Magnetic Hole in the Turbulent Magnetosheath Plasma. *Astrophys. J.* **836**, L27 (2017).

52. Yao, S. T. *et al.* Observations of kinetic-size magnetic holes in the magnetosheath. *J. Geophys. Res. Space Phys.* **122**, 1990–2000 (2017).

53. Yao, S. T. *et al.* Electron Mirror-mode Structure: Magnetospheric Multiscale Observations. *Astrophys. J.* **881**, L31 (2019).

54. Zhang, H. *et al.* Observations of Whistler-mode Waves and Large-amplitude Electrostatic Waves Associated with a Dipolarization Front in the Bursty Bulk Flow. *Astrophys. J.* **933**, 105 (2022).

55. Zhong, Z. H. *et al.* Observations of a Kinetic-Scale Magnetic Hole in a Reconnection Diffusion Region. *Geophys. Res. Lett.* **46**, 6248–6257 (2019).

56. Yao, S. T. *et al.* Waves in Kinetic-Scale Magnetic Dips: MMS Observations in the Magnetosheath. *Geophys. Res. Lett.* **46**, 523–533 (2019).

57. Kitamura, N. *et al.* Observations of the Source Region of Whistler Mode Waves in Magnetosheath Mirror Structures. *J. Geophys. Res. Space Phys.* **125**, (2020).

**Figures**

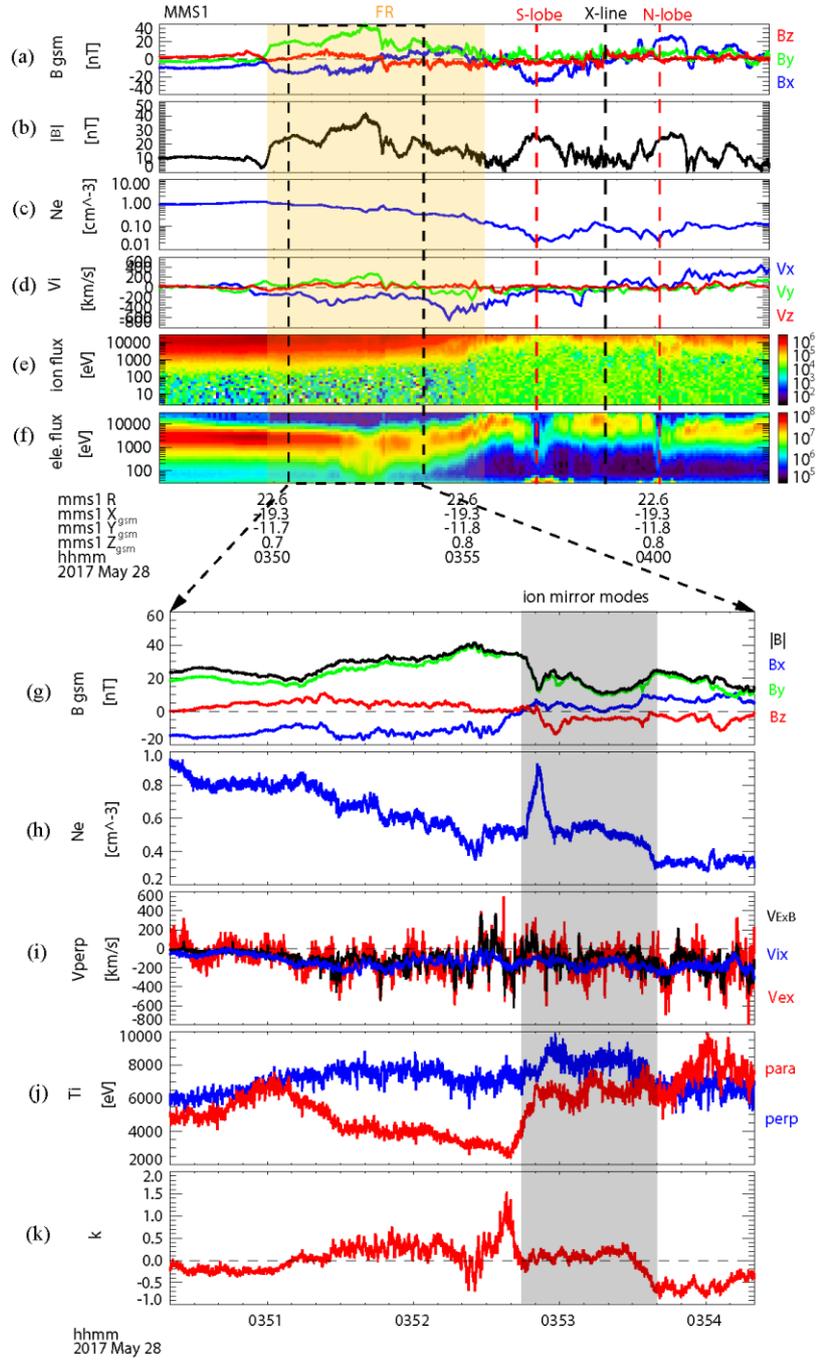

Figure 1. Top: Overview of the magnetotail reconnection observed by MMS1 on May 28, 2017. (a) three components of the magnetic field, (b) total magnetic field, (c) electron density, (d) ion bulk velocity, (e) ion, and (f) electron differential energy flux. Bottom: ion mirror modes observed in the center region of the flux rope. (g) three components and total magnetic field, (h) electron density, (i) **X** component of the perpendicular electron bulk velocity $V_{ex}$, ion bulk velocity $V_{ix}$, and drift velocity $V_{E \times B}$, (j) ion temperature, (k) the parameter $k = \frac{T_{i\perp}}{T_{i\parallel}} - \left(1 + \frac{1}{\beta_\perp}\right)$, the threshold for the ion mirror instability. All vectors are presented in the GSM coordinate system .

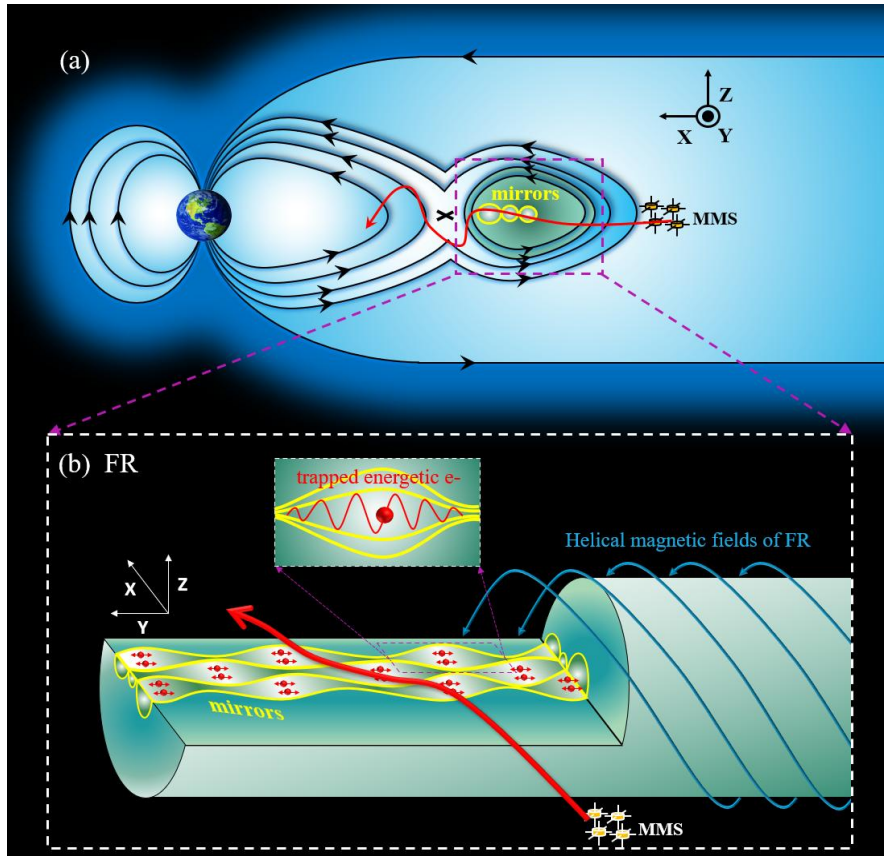

Figure 2. Sketch of the large-scale flux rope and the ion mirror structures. (a) a large-scale FR generated by magnetotail reconnection. The red curve represents the MMS trajectory. (b) three-dimensional sketch of ion mirror structures inside the flux rope and the trapped energetic electrons.

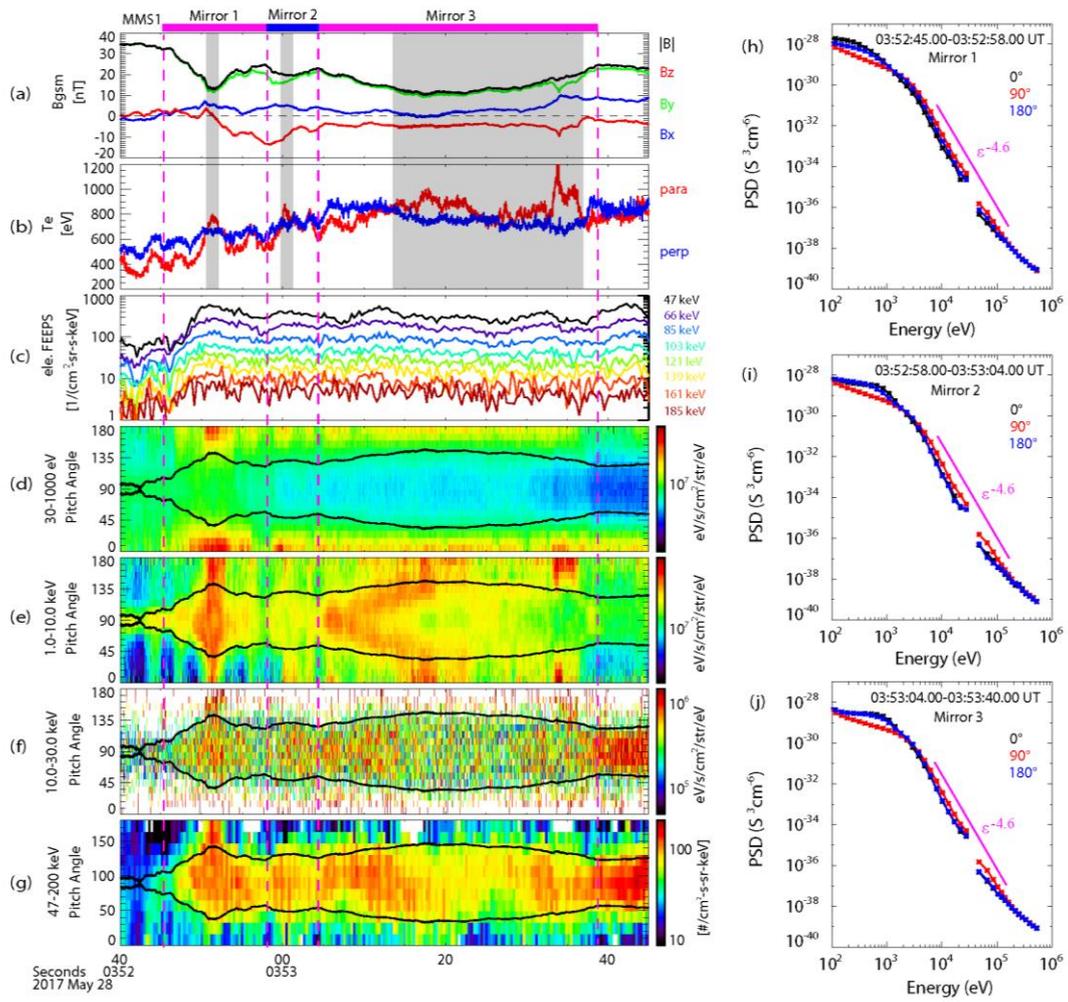

Figure 3. Energetic electrons were observed within the ion mirrors. (a) three components and total magnetic field, (b) electron temperature and $T_{e\parallel} > T_{e\perp}$ region (marked by grey shadows), (c) omni-directional energetic (45-200 keV) electron fluxes measured by FEEPS. pitch angle distribution (PAD) of (d) 30-1,000 eV, (e) 1.0-10.0 keV, (f) 10.0-30.0 keV, and (g) 40-200 keV electrons. The phase space density (PSD) of electrons within (h) mirror 1, (i) mirror 2, and (j) mirror 3. These PSDs show a power-law relation with a slope of −4.6 from about 5 to 200 keV. The magenta vertical dashed lines mark the edges of the mirrors.

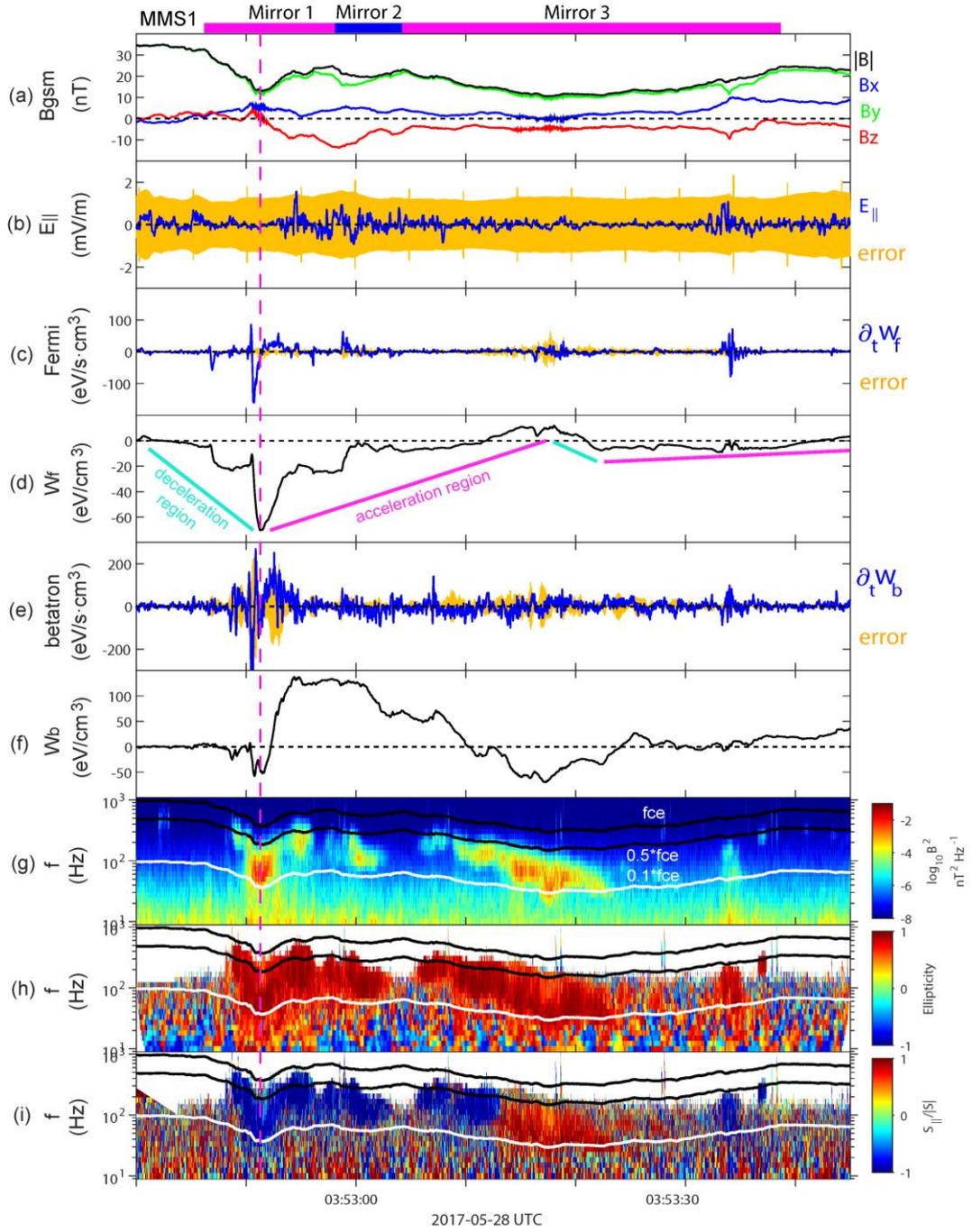

Figure 4. Electron acceleration and whistler waves inside the ion mirror modes. (a) three components and total magnetic field, (b) parallel electric field $E_\parallel$ and its uncertainty, (c) Fermi acceleration rate $\partial_t W_f$ and its uncertainty, (d) integrated $\partial_t W_f$ across this interval $W_f = \int \partial_t W_f dt$, (e) betatron acceleration rate $\partial_t W_b$ and its uncertainty, (f) integrated $\partial_t W_b$ across this interval $W_b = \int \partial_t W_b dt$, (g) B spectrum, (h) ellipticity, and (i) normalized parallel Poynting flux. The magenta vertical dashed line marks the reversal point of $B_z$.